\documentclass[10pt,twocolumn,letterpaper]{article}

\usepackage{cvpr}              

\usepackage{algorithm}
\usepackage{algpseudocode}
\usepackage{amsmath}
\usepackage{multirow}
\usepackage[table]{xcolor}
\usepackage{colortbl}

\definecolor{cvprblue}{rgb}{0.21,0.49,0.74}
\usepackage[pagebackref,breaklinks,colorlinks,allcolors=cvprblue]{hyperref}

\def\paperID{*****} 
\def\confName{CVPR}
\def\confYear{2026}

\title{CIF: A Constrained Inversion Framework for Reliable Message Extraction in Diffusion-Based Generative Steganography}

\author{Yuqi Qian \quad Yun Cao\textsuperscript{*} \quad MeiYang Lv \quad Haocheng Fu\textsuperscript{*}\\
Institute of Information Engineering, Chinese Academy of Sciences, Beijing, China\\
School of Cyber Security, University of Chinese Academy of Sciences, Beijing, China\\
{\tt\small qianyuqi@iie.ac.cn}
}

\begin{document}
\maketitle
\begin{abstract}
Generative image steganography aims to conceal secret information in generated images without arousing suspicion. However, in practical scenarios involving high-capacity embedding or lossy transmission, existing methods still suffer from limited extraction accuracy. The main challenge lies in accurately recovering the secret-embedded latent vectors from stego images.
To address this issue, we propose \textit{CIF}, a constrained inversion framework designed to achieve accurate message extraction. Specifically, CIF reduces dynamic structural errors by enforcing linear consistency in the latent space, meanwhile reduces numerical integration errors by adaptively adjusting the integration order according to local trajectory stability.
Experimental results show that our method reduces latent reconstruction error by more than 35\% and achieves higher message extraction accuracy than existing approaches.
\end{abstract}    
\section{Introduction}
\label{sec:intro}

Steganography aims to hide secret information into ordinary digital media so that the very existence of the message remains undetectable\cite{b:1,Baluja2017}. With the continuous advancement of generative models and the growing presence of AI-generated content (AIGC) on social networks, such content is increasingly becoming a common type of digital media. This trend makes the use of AIGC for covert communication both natural and practical. Generative steganography uses generative models to create images directly conditioned on secret information, embedding messages during generation process. Among these, GAN-based methods\cite{Zhu2018,Zhang2019,Liu2022,Wei2023} usually require an additional decoder network to recover the message, which weakens the coupling between the generator and the extraction process, making it difficult to jointly optimize capacity and fidelity. Normalizing Flow–based approaches\cite{Kingma2018,a:9} are theoretically invertible, but their invertibility constraint limits the expressive capacity of each module, thereby reducing the overall generation quality\cite{Wei2023}. Recently, Diffusion Models\cite{i:1,i:20,a:33,Song2021} have emerged as a promising direction in generative steganography, due to their superior generative fidelity and stable probabilistic modeling\cite{psyduck}.

\begin{figure}[t]
\centering
\includegraphics[width=1.0\columnwidth,trim=50 25 20 50, clip]{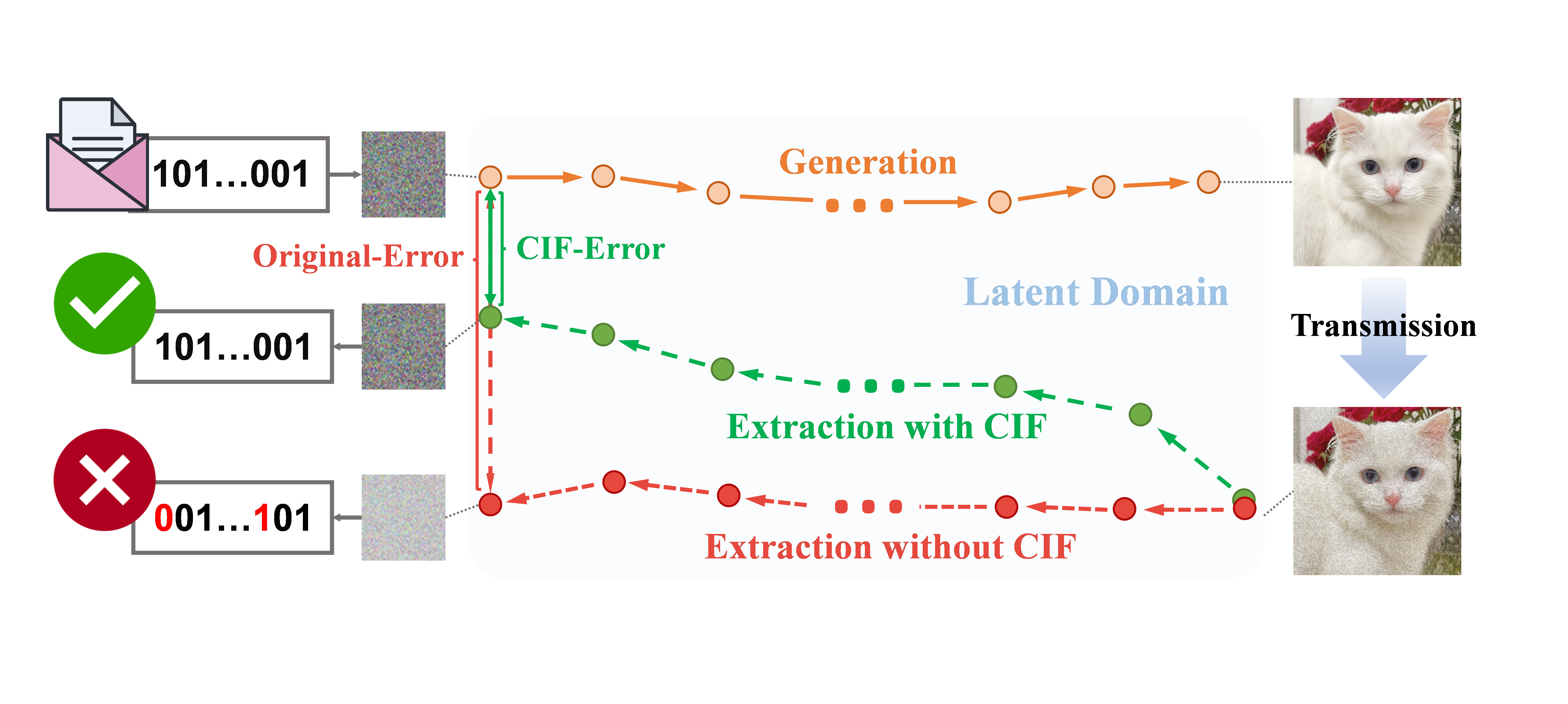} 
\caption{The process of diffusion-based generative image steganography schemes and proposed CIF.}
\label{intro}
\end{figure}

As shown in Figure~\ref{intro}, in a typical diffusion-based steganographic system, the secret message is encoded into the initial latent vectors, then the sender directly generate stego images using this latent vectors via the forward process. During extraction, the receiver reconstructs the latent vectors by performing an inversion of the diffusion process using the model’s predicted velocity field, and then decodes the embedded message. Therefore, the reliability of this reconstruction fundamentally determines the feasibility of generative steganography.
Existing diffusion-based steganography\cite{i:19,a:12,a:13,a:14,Zhu2024,Xu2025,Yu2023,Peng2024} studies primarily concentrate on optimizing the embedding mechanism to improve steganographic performance in terms of capacity, security, and imperceptibility. However, these methods generally fail to constrain the reliability of the inversion process, which is crucial for accurate message extraction. In applications, the reconstruction errors leads to degraded extraction accuracy, especially under high-capacity or distortion conditions.

We conduct a careful study and reveals that the inaccuracy of diffusion inversion mainly arises from two sources. The first is the \textit{dynamic structural error}, caused by the locally linear assumption in reverse integration, which fails to capture the curved and nonlinear nature of the true forward trajectories. The second is the \textit{numerical integration error}, stemming from fixed low-order solvers (e.g., Euler, DDIM) whose truncation errors accumulate across timesteps, while high-order solvers incur substantial computational overhead yet offer negligible accuracy gains in stable regions of latent evolution.

To fix these issues, we propose a constrained inversion framework to achieve an end-to-end inversion process that is geometrically coherent, numerically stable, and computationally efficient. This design substantially enhances the precision of latent vector reconstruction, enabling practical and highly accurate steganographic extraction.
The main contributions of this work are summarized as follows:
\begin{itemize}
\item We propose \textbf{CIF}, a constrained framework that refines diffusion inversion from the perspective of steganographic extraction reliability. CIF improves message recovery accuracy by jointly controlling geometric and numerical errors, ensuring physically consistent and high-fidelity inversion in generative steganography.

\item To reduce dynamic structural errors during extraction, we introduce the Path Consistency Inversion (PCI) module to enforces global directional alignment between forward and reverse diffusion trajectories. We instantiate it using Rectified Flow, which promotes linearized and velocity-matched latent evolution, thereby suppressing curvature-induced deviations and restoring reversible latent dynamics.

\item To reduce numerical integration errors, we introduce the Adaptive Hybrid-Order Solver (AHOS) as a consistency-guided numerical module that adaptively adjusts the integration order according to the local stability of the latent velocity field. By allocating higher-order precision only where instability arises, AHOS achieves high numerical precision while maintaining practical computational efficiency.

\item Extensive experiments validate the effectiveness of the proposed framework. Compared with representative generative steganography methods, our method reduces latent reconstruction error by more than 35\% and achieves the highest message extraction accuracy in practical lossy channels, including JPEG compression, resizing, and other common distortions.

\end{itemize}

\section{Preliminary and Related Work}
\label{sec:preliminary}

\subsection{Diffusion-based Generative Image Steganography}
Diffusion-based generative image steganography utilizes diffusion models to conceal secret data during the image generation process. Existing studies mainly focus on designing effective embedding mechanisms. For example, one approach embeds secret bits into the DCT coefficients of the initial latent noise vector, which preserves the noise distribution and enables nearly lossless message extraction through an invertible diffusion path\cite{a:12}. Another method proposes a dual-key mapping scheme that transforms the message into a Gaussian-distributed latent vector, ensuring statistical indistinguishability from standard noise and enhancing extraction robustness under lossy channels\cite{a:13}. A subsequent work introduces a distribution-consistent embedding mechanism that maps the message into the latent space while maintaining Gaussian characteristics; by offering adjustable embedding modes, it achieves a trade-off between extraction accuracy and payload capacity\cite{a:14}. CRoSS exploits controllable diffusion sampling to jointly enforce cover fidelity, robustness, and security in a unified framework\cite{Yu2023}; and Plug-and-Hide provides a provable and adjustable diffusion steganography mechanism with an explicit trade-off between security and capacity\cite{Zhu2024}. Despite their effectiveness, these methods all rely on precise inversion for accurate message recovery, yet lack dedicated strategies to ensure inversion reliability, leading to suboptimal extraction performance.

\subsection{Forward and Reverse Dynamics in Diffusion Models}
In this paper, we formalize the diffusion dynamics that govern the forward and reverse processes. We define the diffusion dynamics as a continuous mapping that integrates the latent vectors $x_t$ across a time interval $[t_1,t_2]$ under model parameters $\theta$:
\begin{equation}
    x_{t_2} = \Phi_\theta(x_{t_1}; [t_1, t_2])
\end{equation}
where $\Phi_\theta(\cdot; [t_1, t_2])$ represents the integration operator that propagates the latent state along the learned velocity field from time $t_1$ to $t_2$. The forward generation corresponds to $[T,0]$, producing the final image latent $x_0$, while inversion corresponds to $[0,T]$, reconstructing the initial latent vectors $\hat{x}_T$.

In numerical implementations, the forward process is typically discretized as an explicit integration scheme:
\begin{equation}
    x_{t_{k-1}}=x_{t_k}-\upsilon_{\theta}(x_{t_{k}},t_k)\Delta t
\end{equation}
while the reverse inversion proceeds in the opposite temporal direction:
\begin{equation}
    \widehat x_{t_{k+1}}=\widehat x_{t_k}+\upsilon_{\theta}(\widehat x_{t_{k}},t_k)\Delta t
\end{equation}
where $\upsilon(x_t,t)$ denotes the learned velocity field and $\Delta t$ is the integration step size. Here, $t_k$ represents the discrete timestep index in the numerical schedule, and $k \in \{0,1,...T\}$ indexes the successive integration steps along the temporal trajectory.


\section{Methodology}

In this section, we first analyze the underlying causes of inaccuracy in message extraction. Then we introduce our proposed CIF and then elaborate on its two key components: the Path Consistency Inversion (PCI) for geometric alignment and the Adaptive Hybrid-Order Solver (AHOS) for numerical stability.

\subsection{Problem Analysis}
To understand the limitations of existing inversion processes, we analyze the sources of reconstruction inaccuracies in diffusion-based steganography. Our study reveals that these inaccuracies primarily originate from two factors: \textbf{dynamic structural error} and \textbf{numerical integration error}, which jointly affect the accuracy of message extraction.
\subsubsection{Dynamic Structural Error} 

\textbf{Local Linearity Failure} During inversion, the explicit integration scheme specified in Equation (3) is commonly adopted.This formulation implicitly assumes that the latent trajectory remains approximately linear within a small step $\Delta t$, thus enabling a first-order local approximation for backtracking. In practice, however, the forward diffusion trajectory $\gamma _f(t)$ exhibits complex nonlinear curvature structures in the latent space. When the trajectory undergoes sharp turns or high curvature regions, the linear approximation captures only the tangential component while introducing a non-negligible normal deviation, a second-order geometric bias. These local deviations systematically accumulate during multi-step integration, causing the reconstructed trajectory $\gamma _b(t)$ to diverge from the forward one $\gamma _f(t)$. \\
\textbf{Forward–Reverse Asymmetry} The diffusion model is inherently trained to learn the forward generation dynamics (from $x_T \to x_0$), focusing on noise removal or velocity prediction $\upsilon_{\theta}(x_t,t)$. During inversion, one may formally reuse the same function in a reverse-time formulation:
\begin{equation}
    \dot{\gamma}_b(t)=+\upsilon_{\theta}(\gamma_b(t),t)
\end{equation}
yet the model has never been explicitly trained for the reverse dynamics(from $x_0 \to x_T$). This training–inference asymmetry leads to a mismatch in both direction and magnitude, where $\dot{\gamma}_b(t) \neq -\dot{\gamma}_f(t)$. Such trajectory inconsistency accumulates along the integration process, introducing significant deviations in the reconstructed latent vectors.

\subsubsection{Numerical Integration Error}

Inversion is also affected by discretization on a finite time grid. Commonly used solvers such as Euler and DDIM adopt explicit low-order schemes whose local truncation errors are typically $ O(\Delta t^2)$ or $O(\Delta t^3)$. When accumulated over many steps, these local errors introduce a critical global offset along the reconstructed trajectory, which becomes a dominant source of deviation in latent reconstruction:
\begin{equation}
    E_{total}=\sum_{k}^{} \left \| \widehat{x}_{t_k}-x_{t_k}  \right \| 
\end{equation}
High-order solvers are not suitable for diffusion-based steganography, as their substantial computational cost makes them impractical for real extraction scenarios. Moreover, in locally stable regions of the latent dynamics, the inversion has effectively converged, rendering the additional computation of high-order updates unnecessary and offering no meaningful gain in accuracy. These limitations make fixed high-order solvers infeasible for reliable and efficient steganographic inversion.

\begin{figure*}[t]
\centering
\includegraphics[width=1.0\textwidth,trim=20 0 0 20,clip]{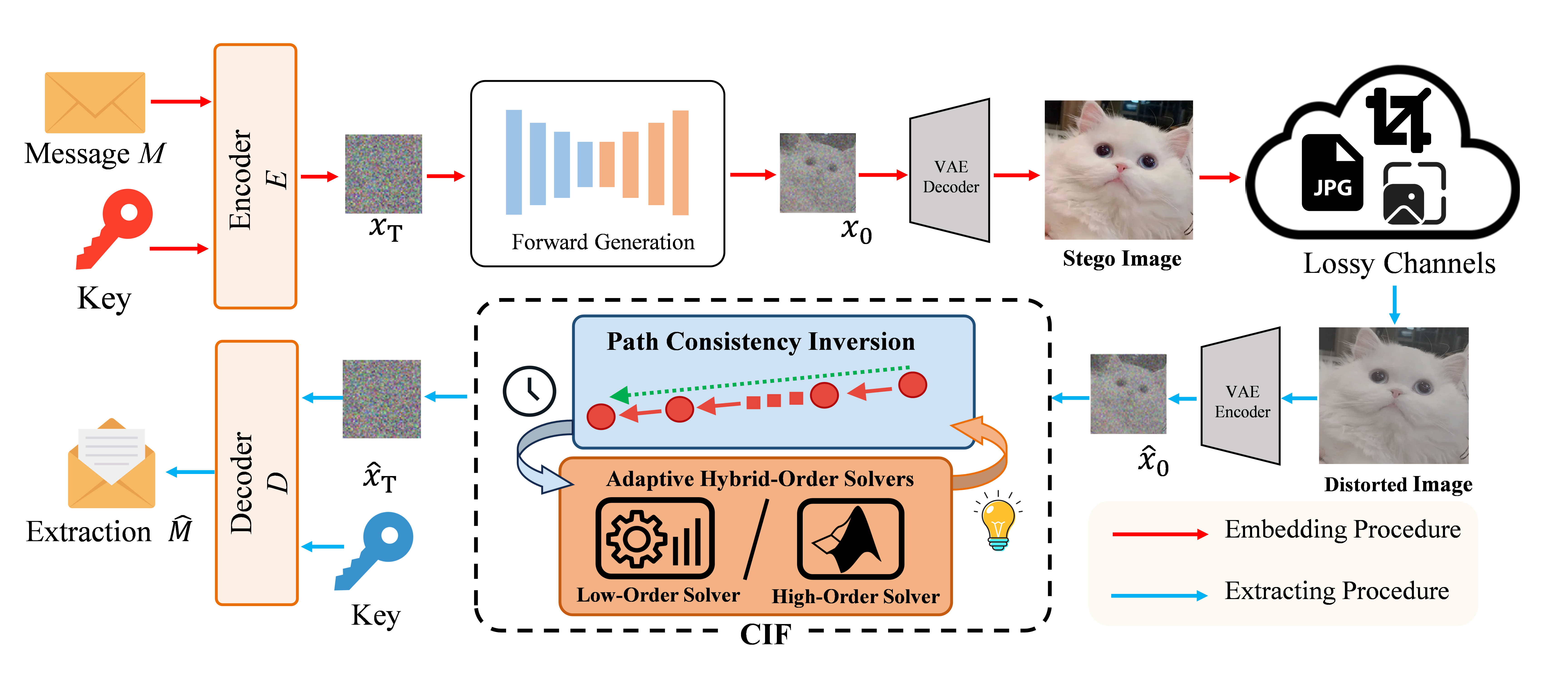} 
\caption{
Architecture of the proposed \textbf{Constrained Inversion Framework (CIF)} combining PCI and AHOS modules. \textbf{Path Consistency Inversion (PCI)} enforces geometric alignment of forward and reverse trajectories,
while the \textbf{Adaptive Hybrid-Order Solver (AHOS)} adaptively regulates numerical precision.
Together they ensure accurate and efficient recovery of hidden information.
}
\label{framework overview}
\end{figure*}


\subsection{The Overview of CIF}

Our framework CIF is designed to suppress the above mentioned inversion errors during extraction. Specifically, it introduces the Path Consistency Inversion (PCI) to mitigate dynamic structural error and the Adaptive Hybrid-Order Solver (AHOS) to reduce numerical integration error. The overall communication process using CIF is summarized as follows.

As illustrated in Figure~\ref{framework overview}, the process begins with the sender encoding a binary secret message $m$ into the initial latent vectors $x_T = E(m)$ through a steganographic encoder $E$. These latent vectors serve as structured noise seeds that drive the forward diffusion process $x_0 = \Phi_\theta(x_T; [T, 0])$, generating the stego image $x_0$ in the pixel domain via the pretrained diffusion model and its VAE decoder. 
At the receiver side, CIF performs the inversion $\hat{x}_T = \Phi_\theta(\hat{x}_0; [0, T])$ by traversing the diffusion dynamics in reverse with two cooperative mechanisms. PCI enforces geometric alignment between the forward and reverse trajectories, ensuring that the reconstructed path remains consistent with the true generative dynamics. Meanwhile, AHOS dynamically adjusts the numerical integration order according to the local stability of the latent velocity field, maintaining low numerical error without unnecessary computation.
Together, these modules enable the reconstructed latent $\hat{x}_T$ to remain within the steganographic decision boundary, ensuring that the extracted message $\hat{m} = D(\hat{x}_T)$ closely matches the original $m$ with high reliability even under perturbations.

\subsection{Path Consistency Inversion}

\subsubsection{Theoretical Formulation}
To constrain the dynamic structural error, we propose the Path Consistency Inversion. PCI enforces a linear trajectory constraint in the latent space, requiring that both the forward generation and reverse inversion processes evolve along a shared near-linear path connecting the initial latent vectors $x_T$ and the final $x_0$. To formalize this principle, we derive the mathematical relationship between the linear trajectory constraint and the achievable inversion accuracy, showing that a sufficiently constrained trajectory directly guarantees precise recovery.

We denote by $u$ the unit vector pointing from $x_T$ to $x_0$, representing the chord direction between the two endpoints,
\begin{equation}
    u=\frac{x_0-x_T}{\left \| x_0-x_T \right \| } 
\end{equation}
PCI constrains the forward and reverse trajectories $\gamma _f (t)$ and $\gamma _b (t)$ to evolve approximately along this direction:
\begin{equation}
    \dot{\gamma _f} (t)=a(t)u,\dot{\gamma _b} (t)=-b(t)u,a(t),b(t)\ge 0
\end{equation}
with boundary conditions $\gamma _f (0)=x_T$, $\gamma _f (T)=x_0$ and $\gamma _b (0)=x_0$, $\gamma _b (T)=\hat{x}_T$. Intuitively, $a(t)$ and $b(t)$ represent the forward and backward progression rates along the same latent line.
When $a(t)$ and $b(t)$ are well matched, the reverse trajectory exactly retraces the forward one, eliminating geometric inconsistency. By integrating the differential relations of both trajectories, the reconstruction error can be expressed as
\begin{equation}
    \widehat{x} _T-x_T=\int_{0}^{T}(\dot{\gamma _b} (t)+\dot{\gamma _f} (t)) dt
    =\int_{0}^{T}(a(t)-b(t))u dt 
\end{equation}
Therefore, the overall reconstruction error is directly determined by the mismatch integral $\int_{0}^{T} |a(t)-b(t)|dt$. A smaller mismatch indicates higher trajectory alignment and leads to more accurate latent reconstruction.
This analysis establishes a clear geometric interpretation: enforcing linear and velocity-matched trajectories in latent space effectively suppresses dynamic structural error, ensuring geometrically consistent and numerically stable inversion.

From a dynamical perspective, the linear trajectory constraint of PCI directly addresses the two core sources of inversion inaccuracy identified earlier. 
First, by enforcing global linearity rather than relying on stepwise local approximations, PCI eliminates the local linearization failure that arises when the diffusion trajectory exhibits high curvature or rapid directional changes. 
The inversion process is no longer limited by locally valid tangential estimates but instead follows a globally consistent latent manifold, thereby preventing curvature-induced accumulation of deviation. Second, by symmetrically constraining the forward and reverse propagation rates $a(t)$ and $b(t)$ along the same direction $u$, PCI restores forward--reverse dynamical symmetry that conventional diffusion models inherently lack. 
This velocity matching ensures that the backward integration retraces the physical evolution of the forward process, suppressing structural asymmetry and closing the reversibility gap. 
Consequently, PCI provides a principled solution to both structural and dynamical sources of inversion error, establishing a geometrically coherent path that guarantees precise recovery.

\subsubsection{Practical Realization}
The theoretical formulation of PCI indicates that an ideal inversion requires a velocity field that evolves linearly along the endpoint chord, maintaining matched forward and reverse dynamics. Rectified Flow (RF) inherently satisfies these requirements. By reparameterizing both the time schedule and the model’s velocity field, RF explicitly minimizes the deviation between instantaneous and global chord directions during training. As a result, it enforces nearly linear, velocity-aligned latent evolution, allowing the inversion process to remain globally coherent rather than locally approximated. From a theoretical standpoint, RF can be viewed as a practical realization of PCI’s geometric and dynamical constraints. Its reparameterized flow formulation ensures that the latent trajectory remains globally linearized, while the forward and reverse velocities are approximately symmetric in both direction and magnitude. This effectively minimizes the mismatch integral $\int_{0}^{T} \left | a(t)-b(t) \right | dt$ in Equation (8). In practice, adopting RF as the backbone inversion model produces curvature-free, dynamically stable, and reversible trajectories, ensuring that the reconstructed latent vectors remain within the steganographic decision boundary and enabling reliable accurate message extraction.

\subsection{Adaptive Hybrid-Order Solver}
While PCI suppresses dynamic structural error, the inversion process still suffers from numerical integration errors that accumulate over discrete timesteps. To overcome this limitation, we introduce the Adaptive Hybrid-Order Solver (AHOS), which dynamically adjusts the integration order according to the local stability of the latent trajectory. Before describing its mechanism, we first present an empirical observation that motivates this design.

\begin{figure}[ht]
\centering
\includegraphics[width=0.5\textwidth,trim=50 0 0 25,clip]{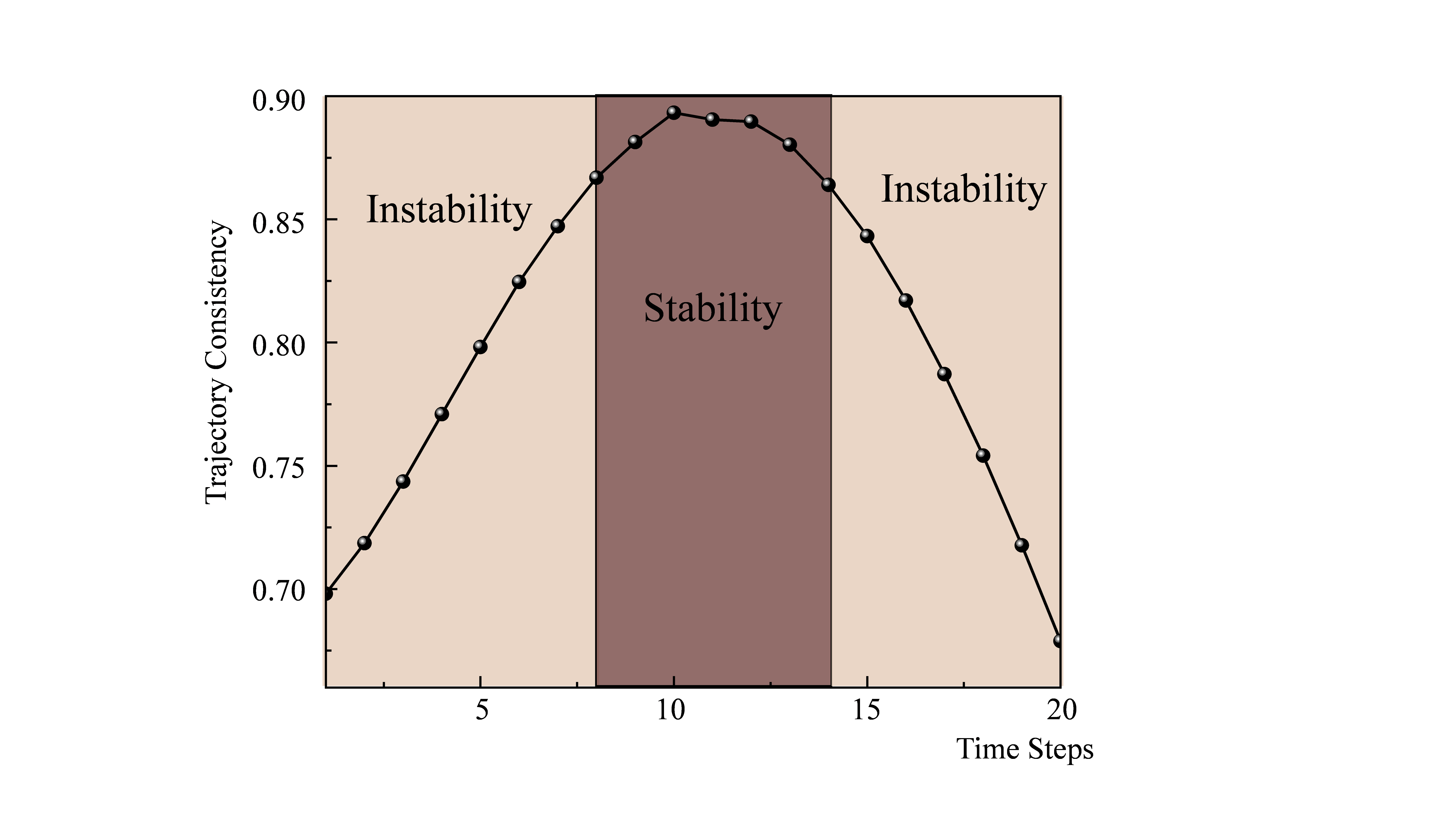} 
\caption{Temporal distribution of trajectory consistency in diffusion inversion.}
\label{cos}
\end{figure}

\subsubsection{Motivation: Temporal Inconsistency in Diffusion Inversion}
To examine how integration errors evolve over time, we analyze the temporal distribution of trajectory consistency between forward generation and reverse inversion. Specifically, we compute the cosine similarity of stepwise velocity directions at corresponding timesteps,
\begin{equation}
    s(t)=\frac{\left \langle \dot{\gamma}_f(t),\dot{\gamma}_b(t) \right \rangle }{\left \| \dot{\gamma}_f(t)\cdot \dot{\gamma}_b(t) \right \| }
\end{equation}
which measures the directional alignment between the two dynamical processes. As illustrated in Figure~\ref{cos}, the trajectory overlap is highest in the mid-range of the diffusion process but degrades significantly at both the early and late timesteps. This nonuniform temporal pattern reveals that inversion instability is not random but systematically concentrated near the endpoints of the diffusion schedule, where the latent dynamics change most rapidly. Such uneven consistency distribution indicates that a fixed-order solver commonly used in diffusion inversion cannot adapt to the changing stability levels over time. Therefore, a dynamic integration strategy is needed to allocate higher numerical precision only in regions where errors are most likely to accumulate.

\subsubsection{Adaptive Integration Mechanism}
Motivated by this observation, AHOS employs a consistency-guided adaptive integration scheme. The key idea is to regulate numerical precision based on the directional stability of the predicted velocity field. Let the model-prediction velocity at step $t_k$ be
\begin{equation}
    \upsilon _k=\upsilon _\theta (\widehat{x} _{t_k},t_k)
\end{equation}
We quantify directional consistency with the previous step via the cosine similarity
\begin{equation}
    s_k=\frac{\left \langle \upsilon _k,\upsilon _{k-1} \right \rangle }{\left \| \upsilon _k \right \|\left \| \upsilon _{k-1} \right \|  } 
\end{equation}
over the already integrated prefix. When the deviation between $s_k$ and $s_{k-1}$ exceeds a predefined threshold $\tau$, indicating a local instability in the latent velocity field, AHOS adaptively switches to a higher-order integrator (e.g., Heun or RK4) to mitigate local truncation errors and maintain numerical consistency. Otherwise, it keeps a fast low-order step (e.g., Euler). This creates a consistency-driven hybrid-order integration that adapts precision to local dynamical stability while preserving efficiency in smooth regions. Unless otherwise specified, the subsequent experiments employ RK4 as the high-order integrator with $\tau=0.05$, which are empirically validated as effective and stable parameter settings. The overall operating procedure of AHOS is illustrated in Algorithm 1.

\begin{algorithm}[ht]
\caption{Adaptive Hybrid-Order Solver}
\begin{algorithmic}[1]
\State \textbf{Input:} Stego image $x_0$, model $v_{\theta}$, time grid $\{t_0, t_1, \dots, t_T\}$
\State \textbf{Output:} Estimated latent vectors $\hat{x}_T$
\State $\hat{x}_{t_0} = x_0$
\State $v_0 = v_{\theta}(\hat{x}_{t_0}, t_0)$
\State $s_k=s_{k-1}=0$
\For{$k = 1, 2, \dots, T$}
    \State $v_k = v_{\theta}(\hat{x}_{t_k}, t_k)$
    \State $s_k = \frac{\langle v_k, v_{k-1} \rangle}{\|v_k\| \|v_{k-1}\|}$
    \If{$\left |s_k -s_{k-1}  \right | > \tau$}
        \State $\hat{x}_{t_{k+1}} = \text{HighOrderStep}(\hat{x}_{t_k}, v_{\theta}, t_k, \Delta t)$
    \Else
        \State $\hat{x}_{t_{k+1}} = \text{LowOrderStep}(\hat{x}_{t_k}, v_{\theta}, t_k, \Delta t)$
    \EndIf
    \State $s_{k-1} = s_k$
\EndFor
\State \Return $\hat{x}_T$
\end{algorithmic}
\end{algorithm}

\begin{table*}[ht]
\centering
\small
\resizebox{\textwidth}{!}{%
\begin{tabular}{l|c|cccc|ccc|ccc|ccc|ccc}
\toprule
\multicolumn{1}{l|}{\multirow{2}{*}{Method}} &
\multicolumn{1}{c|}{Lossless} &
\multicolumn{4}{c|}{Resize} &
\multicolumn{3}{c|}{JPEG Compression} &
\multicolumn{3}{c|}{Median Blur} &
\multicolumn{3}{c|}{Gaussian Blur} &
\multicolumn{3}{c}{Gaussian Noise} \\
\cmidrule(lr){2-2} \cmidrule(lr){3-6} \cmidrule(lr){7-9} \cmidrule(lr){10-12} \cmidrule(lr){13-15} \cmidrule(lr){16-18}
\multicolumn{1}{l|}{} & \multicolumn{1}{c|}{--} & 0.5 & 0.75 & 1.25 & 1.5 & 90 & 70 & 50 & 3$\times$3 & 5$\times$5 & 7$\times$7 & 3$\times$3 & 5$\times$5 & 7$\times$7 & 0.01 & 0.05 & 0.1 \\
\midrule
S2IRT & 81.17 & 77.66 & 79.64 & 80.59 & 80.83 & 79.06 & 76.53 & 75.02 & 78.37 & 74.92 & 72.52 & 79.60 & 78.15 & 76.20 & 80.17 & 76.15 & 73.78 \\
PARIS & 86.90 & 83.71 & 85.68 & 86.46 & 86.65 & 84.99 & 81.76 & 79.20 & 84.45 & 79.94 & 74.89 & 85.66 & 84.31 & 81.98 & 86.03 & 80.80 & 76.25 \\
GRDH & 98.05 & 94.86 & 97.03 & 97.74 & 97.89 & 96.34 & 92.35 & 88.87 & 95.84 & 90.10 & 82.84 & 97.12 & 95.79 & 93.12 & 97.38 & 90.94 & 84.73 \\
Diffusion-Stego & 98.01 & 57.89 & 66.26 & 79.04 & 79.39 & 59.32 & 56.15 & 56.15 & 59.32 & 54.38 & 52.39 & 65.31 & 59.44 & 55.86 & 70.52 & 57.31 & 55.13 \\
\rowcolor{black!15}
CIF & \textbf{99.02} & \textbf{96.35} & \textbf{98.23} & \textbf{98.76} & \textbf{98.85} & \textbf{97.71} & \textbf{95.08} & \textbf{92.21} & \textbf{97.05} & \textbf{91.69} & \textbf{84.53} & \textbf{98.13} & \textbf{96.80} & \textbf{94.00} & \textbf{98.59} & \textbf{93.95} & \textbf{88.23} \\

\bottomrule
\end{tabular}%
}
\caption{Extraction accuracy (\%) comparison of different methods under various distortions. Our method is highlighted with a \colorbox{black!15}{shaded background}, and the best performance in each setting is marked \textbf{in bold.}}
\label{tab:comparison}
\end{table*}

\section{Experiments}
\subsection{Experimental Settings}
We instantiate our framework using the pretrained Rectified Flow (RF-2)\cite{RF} combined with the Adaptive Hybrid-Order Solver (AHOS). Secret messages are embedded via a dual-key noise mapping scheme which achieves balanced performance in terms of embedding capacity and security \cite{a:13}. To ensure generality and reproducibility, we use the public pretrained weights without any finetuning. Unless otherwise noted, we set the latent dimensionality to $L$=4×64×64, image size to $H$=$W$=512, guidance scale $\omega$=1.25 and sampling steps are 20. Text-to-image prompts are taken from the COCO val 2017 captions. All experiments are conducted on NVIDIA RTX L40 GPU using PyTorch 2.3.1 and CUDA 11.8.

\subsection{Evaluation Metrics}
We assess steganographic performance along four axes: extraction accuracy, Hiding capacity, security, and robustness. These metrics jointly reflect reliability for covert communication and resistance to detection. Note that trajectory or inversion–error diagnostics are not included in the main results and will appear later in the ablation/analysis section.\\
\textbf{Extraction Accuracy} We measure accuracy with the bit error rate (BER) using Hamming distance
\begin{equation}
    BER=\frac{dist(M,\hat{M})}{L}=\frac{\sum_{l=1}^{L}1{\{M_l\neq\hat M_l}\} }{L}
\end{equation}
We report the extraction accuracy (Acc):
\begin{equation}
    Acc=1-BER
\end{equation}
\\
\textbf{Hiding Capacity} We measure the embedding capacity by the total number of bits that can be embedded in a single generated image. This metric directly reflects the payload capability of the steganographic system under a fixed image resolution.\\
\textbf{Security} Security refers to an adversary’s ability to distinguish between cover and stego images. In our setting, images generated from random noise are regarded as cover images, while those generated from secret messages are considered stego images. We evaluate security using three state-of-the-art steganalyzers, XuNet \cite{a:21} , YeNet \cite{a:22} , and SiaStegNet \cite{a:23}. The detection performance is measured by the detection error rate $P_E$ defined as
\begin{equation}
    P_E=\frac{P_{FA}+P_{MD}}{2} 
\end{equation}
where $P_{FA}$ denotes the false alarm rate and $P_{MD}$ represents the missed detection rate.
\\
\textbf{Robustness} Assessed by measuring the variation in extraction accuracy under different distortion attacks, including:
\begin{itemize}
\item Resize: Images are resized to 0.5×, 0.75×, 1.25×, and 1.5× their original sizes using bilinear interpolation to simulate resolution changes.

\item JPEG Compression: Evaluated at quality factors of 90, 70, and 50 to represent various degrees of compression degradation.

\item Median Blur: Median filtering with kernel sizes of 3×3, 5×5, and 7×7 to simulate mild, moderate, and severe smoothing distortions.

\item Gaussian Blur: Gaussian filtering with kernel sizes of 3×3, 5×5, and 7×7 to model different levels of blur attacks.

\item Gaussian Noise: Additive Gaussian noise with standard deviations of 0.01, 0.05, and 0.1 to simulate different noise intensities.

\end{itemize}

\subsection{Baseline}
We compare against four state-of-the-art steganographic methods spanning different generative paradigms, each selected for its representative design and competitive overall performance:
\begin{itemize}
\item GRDH\cite{a:13}: A robust image steganography approach that supports zero-shot, text-driven synthesis without additional training or finetuning.
\item Diffusion-Stego\cite{a:14}: A training-free diffusion-based method that embeds information via projection in the generative process.
\item S2IRT\cite{a:9}: A flow-based method that leverages the inherent invertibility of Glow to achieve high capacity and strong robustness.
\item PARIS\cite{a:25}: A GAN-based scheme that provides provable security and channel-aware robustness against practical distortions.
\end{itemize}

\subsection{Experimental Results}

\subsubsection{Extraction Accuracy and Robustness}
As summarized in Table~\ref{tab:comparison}, we conducted a comprehensive evaluation of extraction accuracy under both lossless and common distortion conditions. Our method consistently achieves the highest performance across all scenarios. In particular, it maintains about 96\% accuracy under 0.5× Resize, and around 92\% under JPEG compression with QF = 50, with an even larger margin over competing methods. Although the overall accuracy decreases as the kernel size increases in median or Gaussian blurring, our degradation remains the most gradual. Even under Gaussian noise with $\sigma$ = 0.1, the accuracy still exceeds 88\%. These results highlight the strong cross-type robustness of our approach against geometric, compression, smoothing, and noise perturbations.

\subsubsection{Security}
To evaluate the security of our method, we employed three representative steganalysis networks, namely YeNet, SiaStegNet, and XuNet, to classify cover and stego images. Specifically, we generated 5,000 cover images from random noise and created an equal number of corresponding stego images by embedding secret messages. The entire dataset was systematically split into training, validation, and test sets with a ratio of 4000:500:500.

Table~\ref{tab:steganalysis} summarizes the steganalysis results of all methods. Under our approach, the detection error rate remains close to 50\%, indicating that the steganalyzers fail to distinguish between cover and stego images. This outcome provides strong evidence of the security and stealthiness of our method, demonstrating its effectiveness in resisting state-of-the-art deep steganalysis attacks.

\begin{table}[ht]
\centering
\resizebox{0.9\linewidth}{!}{
\begin{tabular}{l|ccc}
\toprule
Method & YeNet & SiaStegNet & XuNet \\
\midrule
S2IRT             & 0.5000 & 0.4950 & 0.5020 \\
PARIS             & 0.5100 & 0.5085 & 0.4940 \\
GRDH              & 0.5060 & 0.5035 & 0.4910 \\
Diffusion-Stego   & 0.4632 & 0.5523 & 0.4762 \\
\rowcolor{black!15} 
CIF & 0.5012 & 0.4952 & 0.5132 \\
\bottomrule
\end{tabular}
}
\caption{Security comparison of different methods on steganalyzers.}
\label{tab:steganalysis}
\end{table}

\subsubsection{Hiding Capacity}
We further evaluate the embedding capacity of our framework under the standard 512×512 image setting. CIF supports the embedding of 16,384 bits per image, corresponding to a payload of 0.0625 bits per pixel. This configuration maintains a practical balance between payload and imperceptibility while ensuring consistent steganographic reliability. The use of high-resolution images also aligns with real-world social-media distributions \cite{a:33}, improving generation realism and reducing the likelihood of arousing suspicion during transmission.


\subsection{Ablation Study}

We further conduct ablation studies to evaluate the contributions of the two key mechanisms in our framework, Path-Consistent Inversion (PCI) and the Adaptive Hybrid-Order Solver (AHOS), to overall inversion accuracy and efficiency. All experiments are performed under distortion-free conditions. In the ablation settings, the sampling framework based on PCI is replaced with the standard DDIM inversion to remove the path consistency constraint, and the AHOS module is replaced with a fixed first-order Euler solver to disable adaptive integration. To comprehensively assess inversion efficiency, we additionally measure the decoding latency, defined as the average time required to decrypt a single stego image and recover its embedded information. All measurement settings, including hardware and resolution, are kept consistent with those used in the main experiments.

As shown in Table~\ref{tab:ablation}, removing either PCI or AHOS leads to a clear degradation in both accuracy and efficiency. When both modules are disabled, the accuracy drops to 96.23\% and the decoding latency rises to 3345 ms, which indicates that the absence of path consistency and adaptive integration causes error accumulation and unnecessary computation during inversion. Using only PCI improves accuracy to 98.05\% and reduces the latency to 549 ms, showing that enforcing a consistent inversion path can make the decoding process very efficient. Using only AHOS yields a higher accuracy of 98.95\% but the latency remains relatively large at 2138 ms, which suggests that adaptive order selection mainly stabilizes the numerical process but cannot fully remove redundant steps on its own.
With both PCI and AHOS enabled, the full framework reaches the best overall result, achieving 99.02\% extraction accuracy with a moderate latency of 1352 ms. This indicates that the two mechanisms are complementary and that the complete framework attains the highest steganographic fidelity while keeping the computational cost reasonable.

\begin{table}[htbp]
\centering
\resizebox{1.0\linewidth}{!}{
\begin{tabular}{lcc}
\toprule
Method & Accuracy (\%) & Extraction Latency(ms) \\
\midrule
w/o PCI, w/o AHOS & 96.23 &3345\\
w/o PCI & 98.05 & 2138\\
w/o AHOS & 98.95 & 549\\

\midrule

CIF (Full) & 99.02 & 1352\\
\bottomrule
\end{tabular}
}
\caption{Ablation study of Path-Consistent Inversion (PCI) and Adaptive Hybrid-Order Solver (AHOS).}
\label{tab:ablation}
\end{table}

\subsection{Mechanism Validation}
To further elucidate the dynamical origin of the performance improvement brought by the CIF, we design four complementary theoretical metrics that jointly quantify the inversion process from four perspectives: error magnitude, extreme deviation, trajectory directional consistency, and path convergence. Their definitions are as follows:
\begin{itemize}
\item Mean Absolute Error (MAE): Measures the average element-wise deviation between the reconstructed latent vectorss and their original counterparts, reflecting the overall reconstruction bias.

\item Maximum Element Error (Max Error): Records the largest element-wise difference between the inverted and original latent vectorss, representing the upper bound of inversion error. This value partially indicates whether the inversion results remain bounded by the embedding threshold.

\item Angular Consistency (Angle Accumulation): Computed by accumulating the angle between the forward and backward trajectories at each time step, this metric characterizes the directional alignment of their dynamical evolution. A smaller accumulated angle implies that the inversion trajectory follows a more stable and temporally consistent path on the latent manifold.

\item Path Straightness Ratio (Straightness Ratio): Defined as the ratio between the straight-line distance of the trajectory endpoints and its total arc length, this metric evaluates how closely the inversion follows the shortest feasible dynamical path. A value approaching 1 indicates reduced local oscillations and redundant deviations during the inversion process.
\end{itemize}

Together, these four metrics form a comprehensive characterization of the inversion dynamics: MAE and Max Error describe global and extreme error control in the magnitude domain, while Angle Accumulation and Straightness Ratio capture trajectory consistency and directional stability in the geometric domain.

Table~\ref{tab:mechanism} presents the quantitative results of the theoretical metrics under different ablation configurations. 
Compared with the baseline model without PCI and AHOS, the full framework achieves consistent improvements across all four indicators. 
Specifically, the Mean Absolute Error (MAE) decreases from 0.3233 to 0.2050, and the Maximum Element Error (ME) decreases from 2.8457 to 1.7020, 
indicating that CIF significantly reduces both overall and extreme reconstruction deviations in the latent space. 
For trajectory-level metrics, the Angular Consistency (AC) drops sharply from 26.82$^{\circ}$ to 7.14$^{\circ}$, 
demonstrating that CIF enforces much stronger directional alignment between forward and reverse trajectories. 
The Straightness Ratio (SR) rises from 0.85 to 0.88, showing that the reconstructed trajectory becomes smoother and closer to a linear path.  


These results collectively verify that the combination of PCI and AHOS effectively suppresses both structural and numerical errors.
PCI improves the geometric coherence of latent trajectories through path consistency, while AHOS adaptively reduces temporal discretization errors during integration.
Their joint operation yields a stable, reversible, and geometrically consistent inversion process, achieving highly accurate reconstruction of latent vectors and enabling reliable extraction of hidden information.
Notably, the enforced path consistency aligns the physical trajectories of forward and reverse diffusion, ensuring that the inversion process retraces the true generative dynamics rather than an approximate numerical path.
This alignment preserves the physical reversibility of the diffusion process, which is essential for stable and precise message recovery.

\begin{table}[h]
\centering
\resizebox{1.0\linewidth}{!}{
\begin{tabular}{lcccc}
\toprule
Method & MAE & ME & AC & SR \\
\midrule
w/o PCI, w/o AHOS & 0.3233 & 2.8457 & 26.82$^\circ$ & 0.8500 \\
w/o PCI & 0.2073 & 1.7089 & 7.99$^\circ$ & 0.8694 \\
w/o AHOS & 0.2056 & 1.7070 & 7.41$^\circ$ & 0.8783 \\
\midrule
CIF (Full)  & 0.2050 & 1.7020 & 7.14$^\circ$   & 0.8802 \\
\bottomrule
\end{tabular}
}
\caption{Quantitative analysis of inversion consistency metrics for different module configurations.}
\label{tab:mechanism}
\end{table}


\section{Conclusion}
We presented CIF, a constrained inversion framework for diffusion-based generative steganography that enhances extraction accuracy through inversion consistency. By integrating Path Consistency Inversion (PCI) for geometric alignment and Adaptive Hybrid-Order Solver (AHOS) for numerical stability, CIF achieves physically coherent and reversible inversion. Experiments verify that CIF reduces the latent reconstruction error by more than 35\% and achieves the best accuracy among existing approaches.

{
    \small
    \bibliographystyle{ieeenat_fullname}
    \bibliography{main}
}


\end{document}